\documentclass[12pt,a4paper,final]{iopart}
    
\usepackage{iopams}
\usepackage{graphicx}
\usepackage{subcaption}

\expandafter\let\csname equation*\endcsname=\relax 
\expandafter\let\csname endequation*\endcsname=\relax 
\usepackage{physics}
\begin{document}
 
%\title[Effect of classical noise on quantum transport efficiency]{Effect of classical noise on quantum transport efficiency}
\title[Efficient quantum transport in a multi-site system combining classical noise and quantum baths]%
{Efficient quantum transport in a multi-site system combining classical noise and quantum baths}
 
\author{Arzu Kurt}
\address{Department of Physics, Bolu Abant \.{I}zzet Baysal University, 14030-Bolu, Turkey}
\address{Turku Centre for Quantum Physics,
Department of Physics and Astronomy, University of Turku, FI-20014 Turun Yliopisto, Finland}
\ead{arzukurt@ibu.edu.tr}

\author{Matteo A. C. Rossi}
\address{QTF Centre of Excellence, Turku Centre for Quantum Physics,
Department of Physics and Astronomy, University of Turku, FI-20014 Turun Yliopisto, Finland}
\ead{matteo.rossi@utu.fi}

\author[cor1]{Jyrki Piilo}
\address{QTF Centre of Excellence, Turku Centre for Quantum Physics,
Department of Physics and Astronomy, University of Turku, FI-20014 Turun Yliopisto, Finland}
%\eads{\mailto{jyrki.piilo@utu.fi}}
%\mailto{author.three@gmail.com}

\begin{abstract}
We study the population dynamics and quantum transport efficiency of a multi-site dissipative system driven by a random telegraph noise (RTN) by using a variational polaron master equation for both linear chain and ring configurations. By using two different environment descriptions -- RTN only and a thermal bath+RTN --  we show that the presence of the classical noise has a nontrivial role on the quantum transport. We observe that there exists large areas of parameter space where the combined bath+RTN influence is clearly beneficial for populating the target state of the transport, and  for average trapping time and transport efficiency when accounting for the presence of the reaction center via the use of the sink. This result holds for both of the considered intra-site coupling configurations including a chain and ring.
In general, our formalism and achieved results provide a platform for engineering and characterizing efficient quantum transport in multi-site systems 
both for realistic environments and engineered systems.

\end{abstract}
%\submitto{\NJP}
%\pacs{}
\vspace{2pc}
%\maketitle

%-------------------------
\section{Introduction}
Recent research has shown that quantum transport efficiency can be enhanced by
the help of environmental noise, an effect known as environment-assisted quantum
transport (ENAQT) \cite{Rebentrost2009b} or dephasing-assisted transport
\cite{Plenio2008,Chin2010}. The phenomenon has received particular attention in
light-harvesting systems, such as photosynthetic complexes, but is also relevant in engineered systems, e.g., superconducting circuits and trapped-ion systems.

The  environment of
the light-harvesting protein has been modeled in different ways, including
random telegraph noise (RTN)
~\cite{Hoyer2010,Localization2011,Nesterov2013,Nesterov2015,Berman2016,Gurvitz2017a,Nesterov2018,Trautmann2018,Gurvitz2019},
the Haken-Strobl-Reineker (HSR) model
~\cite{Wu2010,Vlaming2012,Nesterov2013b,Wu2013,Uchiyama2017a} and collections of
harmonic oscillators
\cite{Plenio2008,Mohseni2008,Rebentrost2009,Cao2009,Caruso2009,Fassioli2010,Chin2010,Montiel2014,Mohseni2014}.
Earlier results show that the quantum transport efficiency may be enhanced for
certain values of the parameters of the noise, such as dephasing
rate~\cite{Mohseni2008,Rebentrost2009b,
Rebentrost2009,Cao2009,Caruso2009,Chin2010,Hoyer2010,Fassioli2010,Localization2011,Montiel2014,Trautmann2018},
noise
amplitude~\cite{Nesterov2013,Nesterov2013b,Wu2013,Berman2016,Gurvitz2017a},
reorganization
energy~\cite{Mohseni2008,Rebentrost2009,Wu2010,Localization2011,Mohseni2014},
and noise
correlations~\cite{Plenio2008,Vlaming2012,Nesterov2015,Uchiyama2017a,Dutta2017}.

For instance, Nesterov \emph{et al.}~\cite{Nesterov2015} have studied the dependence of efficient energy transfer (EET) on the correlation properties of the random fluctuations of the protein environment by modeling those fluctuations by RTN for a donor-acceptor system (i.e., a two-level system). They found that in case of strong-electronic coupling, the independent noise fluctuations on the site energies may be more effective in helping EET than collective noise. 

The effects of RTN on two-level systems are also studied in the field of quantum information \cite{Galperin2006,Rossi2016,Benedetti2013a}, with particular attention to solid-state devices \cite{Paladino2014}.
Uchiyama \emph{et al.}~\cite{Uchiyama2017a} have analyzed the effect of spatial and temporal correlations on  EET in a multi-site model by using a Ornstein-Uhlenbeck noise process to describe the environment, and observe that negative spatial correlation of the noise is the most effective in helping the EET. The effect of RTN on transport via continuous-time quantum walks has been studied on lattices \cite{Benedetti2016,Benedetti2018}, also in presence of spatial correlations \cite{Rossi2017a}. Its effect on the non-Markovianity of the dynamics of the spin-boson model has also been considered \cite{Kurt2018}.

Besides theoretical studies, ENAQT has been reported in experiments with superconducting quantum circuits~\cite{Potocnik2018}, photonic setups~\cite{Biggerstaff2016}, classical oscillators~\cite{DeLeon-Montiel2015} and trapped-ions~\cite{Maier2018}. All these works show that in certain parameter regimes the efficiency of the energy transfer is enhanced by environmental noise.

The assumption that the environment can be modeled as a collection of independent harmonic oscillators might not be adequate for various situations where there exist different types of motions that can not be handled by a harmonic approximation. For example, in light harvesting systems such as the Fenna-Matthews-Olson (FMO) complexes, the transfer of exciton might be affected by large amplitude modes, e.g., motion of the protein scaffold. Besides, the concept of the so-called real environment as both a quantum mechanical thermal bath and a classical noise source representing anharmonic effects for the multi-site model in the open quantum system is lacking~\cite{Jang2018} and the important question remains: can one observe ENAQT-like phenomena in an environment treatment consisting of both a stochastic noise and quantum thermal bath?
 
To answer this question, we study numerically an exact noisy multi-site spin-boson model in the variational polaron framework \cite{Pollock2013} deriving the corresponding master equation, and also add RTN on the site energies. 
We then study the transport efficiency of a single electronic excitation, and compare the effect of classical and quantum noise, and their interplay. Our results show the presence of a nontrivial interplay between the quantum thermal bath and the classical noise -- displaying enhanced transport efficiency. The used formalism and results go beyond a recent study~\cite{Uchiyama2017a} which demonstrated the improved transport efficiency when using (anti)correlated Ornstein-Uhlenbeck noise only.
Our study does not necessarily model a specific real system, but considers ingredients that are in general plausible in the different scenarios described above, in order to shed light on the ENAQT phenomenon and to provide a platform for engineering and characterizing efficient quantum transport in multi-site systems both for realistic environments and engineered systems.

The article is organised in the following way. In Section~\ref{sec:model} we discuss the model and derive the variational master equation. Then 
in Sec.~\ref{sec:results} we present the results of the study, by first looking at the dynamics of the system followed by detailed results on the efficiency of the transport and exciton trapping time. Finally, we conclude with discussion and outlook in Sec. \ref{sec:conclusion}.

\section{Model}
\label{sec:model}

\begin{figure}[t]
   \centering
   \begin{subfigure}[t]{0.45\linewidth}
   \includegraphics[width=\linewidth]{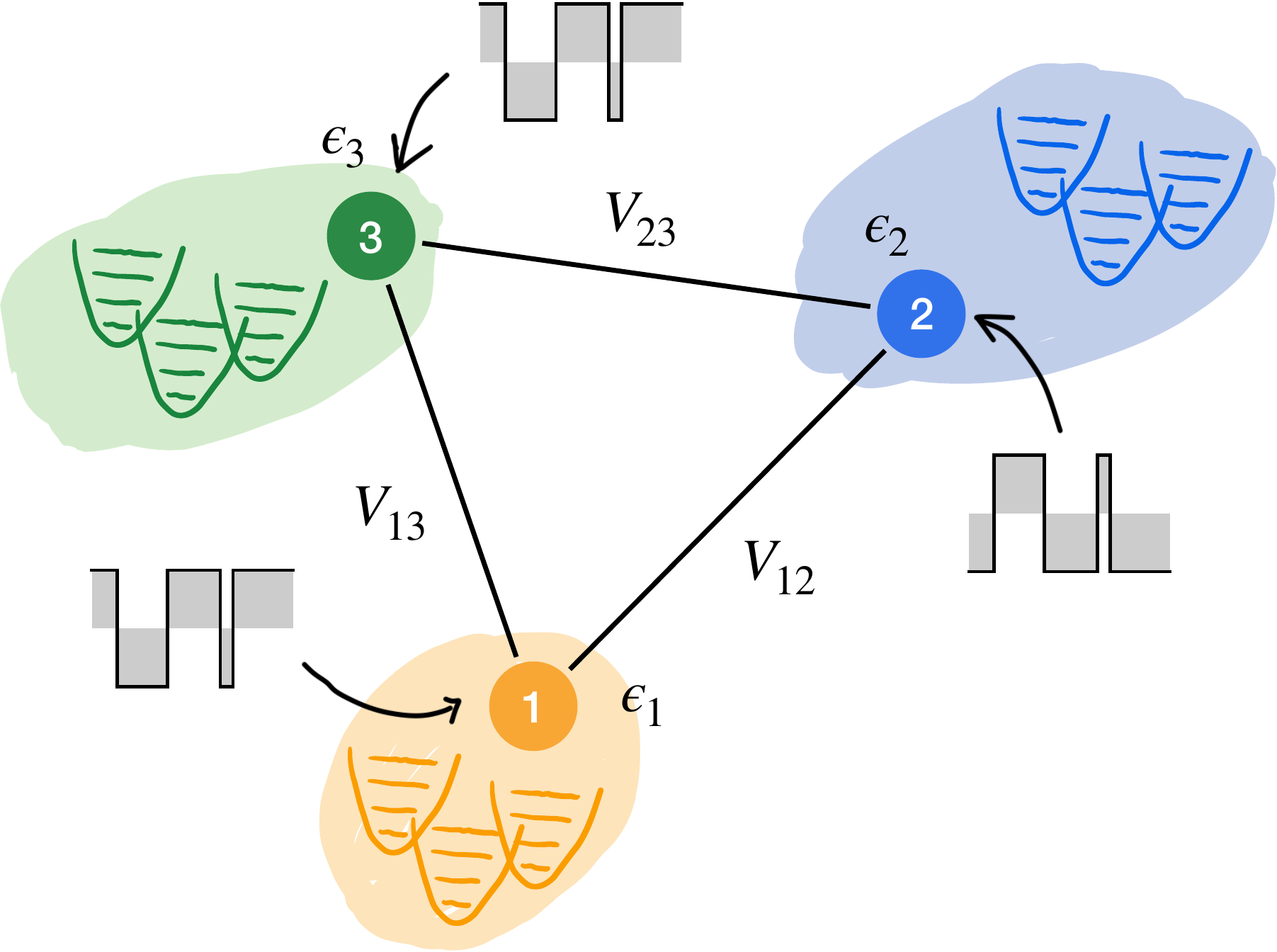}
   \caption{}
   \end{subfigure}
      \centering
   \begin{subfigure}[t]{0.5\linewidth}
      \includegraphics[width=\linewidth]{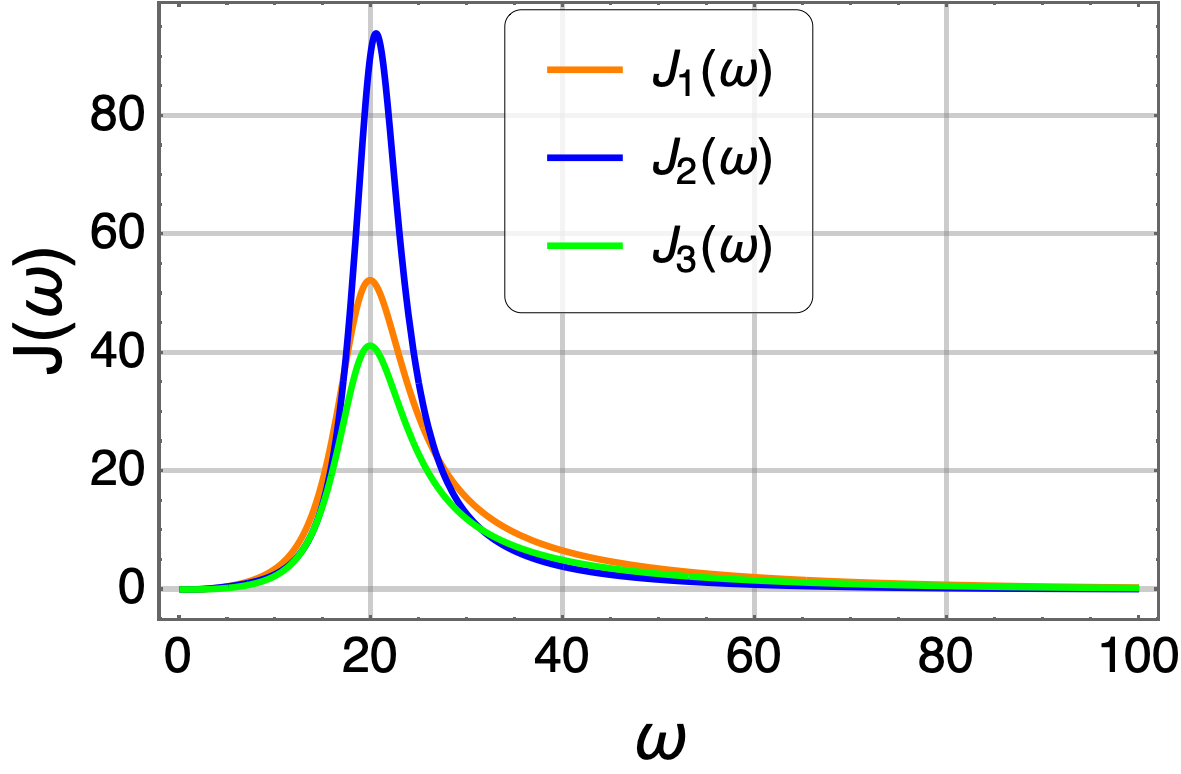}
   \caption{}
   \end{subfigure}
   \caption{(a) A schematic representation of the system under study. The three sites, with energies $\epsilon_i$, are coupled with strengths $V_{ij}$. In the chain configuration (not depicted here), $V_{13} = 0$. Each node is interacting with a thermal bath of harmonic oscillators as described in Eq. \eqref{eq:ham}, and its energy is perturbed by RTN. The noises affecting sites 1 and 3 are completely correlated, and they are completely anti-correlated with respect to the noise affecting site 2. The spectral densities characterizing the baths, Eqs. \eqref{eq:spectJcom} are shown in (b).}
   \label{fig:model}
\end{figure}

We consider a noisy multi-site spin-boson model where each site interacts with a
separate thermal environment, and the site energy is modulated by RTN, as
represented in Fig. \ref{fig:model}a. The
total Hamiltonian of the system is given by

\begin{equation}\label{eq:ham}
    H=H_S(t)+H_E+H_I,
\end{equation}
where $H_S(t)$ is the (time-dependent) Hamiltonian for the spin system, $H_E$ is the Hamiltonian for the bosonic environment and $H_I$ is the interaction part:
\begin{eqnarray*}
H_S(t)&=&\sum_{n}\epsilon_n(t)\dyad{n}+\sum_{n\neq m}V_{nm}\dyad{n}{m},\\
H_E&=&\sum_{n,k}\omega_{n,k}\,b_{n,k}^{\dagger}b_{n,k},\\
H_I&=&\sum_{n,k}\dyad{n}(g_{n,k}b_{n,k}^{\dagger}+g_{n,k}^{*}b_{n,k}).
\end{eqnarray*}

Here $b_{n,k}(b_{n,k}^{\dagger})$ is the annihilation (creation) operator for
the $k$th oscillator of the $n$th site whose state is described by $|n\rangle$,
$V_{nm}$ is the electronic coupling between $n$th and $m$th sites, and $g_{n,k}$
is the interaction strength between the exciton on the $n$th site and $k$th
harmonic oscillator of its bath. We assume that the energy of each site
fluctuates with RTN, i.e.,
$\epsilon_{n}(t)=\epsilon_{n0}+\Omega_n\alpha_n(t)$, where $\epsilon_{n0}$ is the
bare site energy of $n$th site and $\Omega_{n}$ is the noise amplitude at
$n$th site. The RTN is a stochastic process that describes a bistable fluctuator
that jumps between two values $\alpha = \pm 1$ with a certain switching rate
$\nu$. It
is characterized by zero average, $ \langle\alpha_n(t)\rangle=0$, and an exponentially decaying auto-correlation function $\langle\alpha_n(t)\alpha_n(t')\rangle=e^{-\nu|t-t'|}$, where the correlation time of the noise is $\tau_{c}=1/\nu$. 

The bosonic environment of the system is treated as a collection of non-interacting harmonic oscillators. The system-bath interaction is characterized by the spectral density, defined by  $J_n(\omega)=\sum_{k}|g_{n,k}|^{2}\delta(\omega-\omega_k)$. In the continuum limit, the strength of the coupling of the system to the environment is measured by site specific reorganization energy $E^{r}_n=\int_{0}^{\infty}d\omega\,J_n(\omega)/\omega$. For the present study, we consider a spectral density $J_{com}(\omega)$ consisting of two parts. The first part defines the broad range background modes (overdamped) \cite{Kell2013}, while the second one describes a discrete underdamped vibrational \cite{Roden2012} mode which interacts with an Ohmic environment with cut-off frequency $\Lambda$:
\begin{eqnarray}
\label{eq:spectJcom}
    J_{com}(\omega)&=&J_{bg}(\omega)+J_{vib}(\omega),\\
    J_{bg}(\omega)&=&\sqrt{\frac{\pi}{2}}\frac{S \omega}{\sigma}\exp\left[-\frac{1}{2}\left(\frac{\log{[\omega/\omega_ c]}}{\sigma}\right)^2\right]\label{eq:spectJbg}\nonumber,\\
    J_{vib}(\omega)&=&X\omega^{2}\frac{J_{ohm}(\omega)}{(\omega-g(\omega))^{2}+J_{ohm}(\omega)^{2}}\label{eq:spectJvib}\nonumber,\\
    J_{ohm}(\omega)&=&\xi\,\omega\,e^{-\omega/\Lambda},\quad g(\omega)=\zeta-\xi\,\frac{\Lambda}{\pi}+\frac{1}{\pi}J_{ohm}(\omega)\mathrm{Ei}[\omega/\Lambda]\nonumber.
\end{eqnarray}
Here, $\mathrm{Ei}[\omega/\Lambda]$ is the exponential integral function while
parameters $S$ and $X$ determine the peak magnitude of $J_{bg}(\omega)$ and
$J_{vib}(\omega)$, respectively. Moreover, $\xi$ and $\Lambda$ act as damping
factors for the discrete oscillator which corresponds to underdamped mode,
whilst $\zeta$ governs not only the position of the underdamped mode, but also
the magnitude of $J_{vib}(\omega)$ in the present problem. Each site is assumed
to have its own independent environment which is described by site specific
spectral function parameters in Eq.~(\ref{eq:spectJcom}). Figure \ref{fig:model}b shows
the spectral densities of the quantum baths.

When the ratio between the electronic coupling $V_{nm}$ and the bath reorganization energy $E^{r}_n$ is small, the use of Redfield master equation is adequate \cite{Redfiled1957,Breuer2002}, while the full polaron master equation \cite{Jackson1983} is useful when the ratio is large. In many photosynthetic systems, such as the Fenna-Matthews-Olsen (FMO) complex, the convenient coupling regime corresponds to the intermediate regime  \cite{Grondelle2006}. In order to describe the dynamics of the reduced density matrix of the current system, we adopt the variational polaron master equation for the multi-site spin-boson model derived by Pollock \emph{et al.} \cite{Pollock2013}. This is based on using the variational polaron transform

\begin{equation}
G=\sum_{n,k}\dyad{n}\omega_{n,k}^{-1}\left(f_{n,k}b_{n,k}^{\dagger}-f_{n,k}^{*}b_{n,k}\right),
\end{equation} 
which transforms the total Hamiltonian as $\tilde{H}=e^{G}He^{-G}$ to 
\begin{eqnarray}
\tilde{H}&=&\tilde{H}_{0}+\tilde{H}_{I},\quad \tilde{H}_{0}=\tilde{H}_{S}+\tilde{H}_{E},\quad \mathrm{and}\quad \tilde{H}_{I}=\tilde{H}_{L}+\tilde{H}_{D}\label{eq:varPolHam},\\
\tilde{H}_S&=&\sum_{n}(\epsilon_n(t)+R_n)\dyad{n}+\sum_{n\neq m}V_{nm}\,\mathcal{B}_{n}\,\mathcal{B}_{m}\dyad{n}{m}, \quad \tilde{H}_{E}=H_{E}\nonumber,\\
\tilde{H}_L&=&\sum_{n,k}\dyad{n}\left((g_{n,k}-f_{n,k})b_{n,k}^{\dagger}+(g_{n,k}-f_{n,k})^{*}b_{n,k}\right)\nonumber,\\
\tilde{H}_D&=&\sum_{n\neq m}V_{nm}\dyad{n}{m}B_{nm}\nonumber.
\end{eqnarray}

In the transformed interaction Hamiltonian $\tilde{H}_{I}$, we have two types of interaction terms; $\tilde{H}_{L}$ is similar to the weak interaction term, while $\tilde{H}_{D}$ is the full polaron term. $B_{nm}$ is the environmental displacement operator

\begin{equation}
B_{nm}=e^{\sum_{n,k}\left(\frac{f_{n,k}}{\omega_{n,k}}\,b_{n,k}^{\dagger}-\frac{f_{n,k}}{\omega_{n,k}}^{*}b_{n,k}\right)-\sum_{m,k}\left(\frac{f_{m,k}}{\omega_{m,k}}\,b_{m,k}^{\dagger}-\frac{f_{m,k}}{\omega_{m,k}}^{*}b_{m,k}\right)}-\mathcal{B}_n\,\mathcal{B}_m,
\end{equation}
which includes the expectation values $\mathcal{B}_{n}$, defined as

\begin{equation}
\mathcal{B}_{n} =\exp\left({-\frac{1}{2}\sum_{k}
\frac{|f_{n,k}|^2}{\omega_{n,k}^2}\coth\left[\beta\omega_{n,k}/2\right]}\right).
\end{equation}

The electronic coupling $V_{nm}$ and site energy $\epsilon_{n}(t)$ are renormalized respectively by $\mathcal{B}_{n}$ and $R_{n}$, the latter being defined as:
\begin{eqnarray*}
R_{n}&=&\sum_{k}\left(\frac{|f_{n,k}|^2}{\omega_{n,k}}-2\Re[f_{n,k}g_{n,k}^{*}]\right).
\end{eqnarray*}
In the variational polaron transformation, $f_{n,k}\neq g_{n,k}$, and $f_{n,k}$ are left as free optimization parameters which are determined numerically by minimizing the contribution of $\tilde{H}_{I}$ to the free energy~\cite{Pollock2013}. 

The interaction Hamiltonian $\tilde{H}_{I}$ in the variational polaron frame is assumed to be $\tilde{H}_{I}=\sum_{i,j=1}^{N^2}S_i\otimes E_i$, where $N$ is the number of sites, $S_{i}$ and $E_i$ are the system and environment operators, respectively. One can use the projection operator method to obtain a master equation for the system density matrix $\tilde{\rho}_{S}(t)=\mathrm{Tr}_{E}[\tilde{\rho}(t)]$ with the Hamiltonian in Eq.~(\ref{eq:varPolHam}) in the Schr\"{o}dinger picture as \cite{Pollock2013}:

\begin{align}
\label{eq:master}
    \frac{\partial \tilde{\rho}_{S}(t)}{\partial t} = & -i\left[\tilde{H}_{S}(t),\tilde{\rho}_{S}(t)\right]-i\left[\tilde{H}_{trap},\tilde{\rho}_{S}(t)\right]\notag\\
    &-\sum_{i,j=1}^{N^2}\int_{0}^{t}ds\langle E_{i}(s)E_{j}(0)\rangle\left(\{S_i\,S_j(s)\,\tilde{\rho}_{S}(t)-S_j(s)\tilde{\rho}_{S}(t)S_i\}+ hc\right).
\end{align}
    
The term $\tilde{H}_{trap}=i\kappa\,\dyad{n}$ is the trap (or sink)
Hamiltonian~\cite{Rebentrost2009b,Fassioli2010,Montiel2014} --  an
anti-Hermitian operator that is used to describe the excitation leaving the
system permanently. Below, for the studies of population dynamics, this part is
excluded, while it is included for transport efficiency and trapping time
studies. In the latter case, the trapping is assumed to occur only on site $n$
with rate $\kappa$. Above, $\langle E_{i}(s)E_{j}(0)\rangle$ are the bath
correlation functions having four non-zero types [see Eqs.(17)-(21) in
\cite{Pollock2013}]. $S_j(t)=U(t)S_j U^{\dagger}(t)$ is the $j^{th}$ system
operator in the interaction Hamiltonian at time $t$ which is obtained by using
the time evolution operator
$U(t)=\mathcal{T}\exp{\left(-i\int_{0}^{t}H_{S}(t')dt'\right)}$ where
$\mathcal{T}$ is the time-ordering operator. Since
$\left[H_S(t),H_S(t')\right]\ne 0$ due to the presence of RTN on the site
energies, it is not possible to derive a simple expression for $U(t)$ for the
given system Hamiltonian. 

We adopt the following procedure to calculate $U(t)$
for each RTN noise realization. 
Let the state flipping times of the RTN be $\{0,t_1,t_2,\ldots,t_F\}$ in the time interval $[0,t_{F}]$. Since the magnitude of $\alpha(t)$ is either $1$ or $-1$ in between any two of those flipping times, $H_{S}(t)$ is constant in that sub-internal and $U(t)$ can be easily calculated as $\exp(-i H_S\,t)$.  The time evolution operator in the interval $[0,t_1]$ would be either $U_{+}(t)$ or $U_{-}(t)$ depending on whether $\alpha(t)$ is in state $+1$ or $-1$. In general, if $t_n<t<t_{n+1}$, the time evolution operator $U(t)$ would be equal to $U_{+}(t-t_n)\,U_{-}(t_n-t_{n-1})\,U_{+}(t_{n-1}-t_{n-2})\ldots\,U_{+}(t_{2}-t_1)\,U_{-}(t_1)$ if the noise starts from negative values and the number of noise flips in the interval $(0,t_{n+1}]$ is odd.

The master equation \eqref{eq:master} is derived for an arbitrary number of sites. In the following, however, we will focus on a three-site system, in different network configurations, to study the population dynamics, transport efficiency, and exciton trapping times.

%Note to myself : noise averaging of efficiency is performed by integrating the noise averaged populations as outlined in Ref.\cite{Kassal2013}.  described by non-Markovian master equation entering into both spatial and temporal correlations due to the external noise
\begin{table}[!tbh]
\begin{tabular}{p{5.9cm}p{1.2cm}p{1.2cm}p{1.2cm}p{1.2cm}}
\br
\multicolumn{2}{c}{Quantities} & \multicolumn{3}{c}{Sites} \\
& &$1$&$2$&$3$\\ 
\mr
Site energy & $\epsilon_i$&$2v$  &$v$ &0 \\ 
Electronic coupling & $V_i$&$v$  &$v$&0 or $v$ \\ 
Tunneling renormalization& $B_i$&0.87 &0.46 & 0.69\\  
Site energy shift& $R_i$&-22.47 &-38.30 &-24.42\\
Reorganization energy& $E^{r}_i$&33.82 &38.77 &25.94\\ 
Background cut-off frequency& $\omega_{c,i}$&$v$&$2v/3$ &$v$\\ 
Standard deviation& $\sigma_{i}$& 0.7&0.7 &0.7\\  
Peak amplitude factor& $S_{i}$& 0.06&0.04 &0.02\\  
Huang-Rhys factor& $X_{i}$& 0.5& 0.6& 0.4\\  
Damping factor& $\xi_{i}$&0.5 &0.5 &0.5\\ 
Cutoff frequency for Ohmic bath& $\Lambda_{i}$&$2v$ &$3v/2$ &$2v$\\  
Center frequency of $J_{vib}(\omega)$& $\zeta_{i}$&$2v$ &$2v$&$2v$\\ \br
\end{tabular}
\caption{Hamiltonian parameters, the calculated renormalization 
constants of the 3-site system, Eq.~\eqref{eq:varPolHam}, and the parameters of the site specific bath spectral functions, Eq.~\eqref{eq:spectJcom}. $V_{1,2,3}$ correspond to $V_{12}$,$V_{23}$ and $V_{13}$. Also, $V_{13}$ has two values (0 or $v$) depending on the type of coupling configurations. $B_i$, $\sigma_i$, $\xi_i$, $S_i$ and $X_i$ are dimensionless, while all the other parameters are in ps$^{-1}$. Note that $\omega_ {c,i}$, $\sigma_i$, $S_i$ are the parameters of $J_{bg}(\omega)$, while $X_i$, $\xi_i$, $\Lambda_i$ and $\zeta_i$ describe the discrete mode. $v$ is set to 10 for the calculations in the present study. }
\label{tab:par1}
\end{table}

%%%%%%%%%%%%%%%%%%%%%%%%%
\section{Results}
\label{sec:results}
\subsection{The dynamics and steady-state behavior in three-site system}
\label{sub:steady-state}
In this section, we study the dynamics of the system and its steady state in the absence of a sink ($\tilde H_{trap} = 0$). We calculate the time evolution of the system numerically by using the ensemble averaging method based on the average over the noise realizations for a system in which site energy differences are comparable with the coherence electronic site couplings at low temperature ($k_{B}T=1$ps$^{-1}$). The system-bath couplings are assumed to be site dependent, with reorganization energies that are also comparable to the electronic couplings. Although it is customarily assumed that the ratio between the reorganization energy ($E^{r}_i$) and the electronic coupling ($V_{i}$) describes the strength of the system-bath coupling~\cite{Ishizaki2010}, this is not straightforward for the system where each site has a spectral density with different interaction constant. In the present framework, the system-bath interaction strength can be gauged by analyzing the tunneling renormalization factor $B_i$, polaron shift $R_i$, and the site-dependent variational transform functions. As the system-bath coupling increases, $B_i$ tends to zero, while $R_i$ approaches negative of the reorganization energy. 

The system Hamiltonian and the spectral parameters of the thermal environment are given in Table~\ref{tab:par1}. As can be seen from the table, site 2 has the strongest bath coupling, while the bath couplings of site 1 and site 3 are intermediate based on values of $B_i$ and comparison between $R_i$ and $E^{r}_i$. 

We carried out a detailed study of uncorrelated and anti-correlated RTN affecting the three sites. Uncorrelated noise proves to be detrimental for the excitonic transport (see also \cite{Nesterov2015,Uchiyama2017a}, hence in this paper we present the results obtained by a collective noise model. We assume that each site energy level is fluctuating due to the effect of collective random telegraph noise with the form $\{1,-1,1\}\Omega$: this means that the noise at site 2 is anti-correlated with those at sites 1 and 3, which are fully correlated. The results presented below were obtained by averaging the solutions of the master equation~(\ref{eq:master}) over 100 realizations of the RTN process. The convergence of the averages as function of number of noise realizations has been checked carefully.

\begin{figure}[t]
    \centering
    \begin{subfigure}[t]{0.4\linewidth}
    \includegraphics[width=\linewidth]{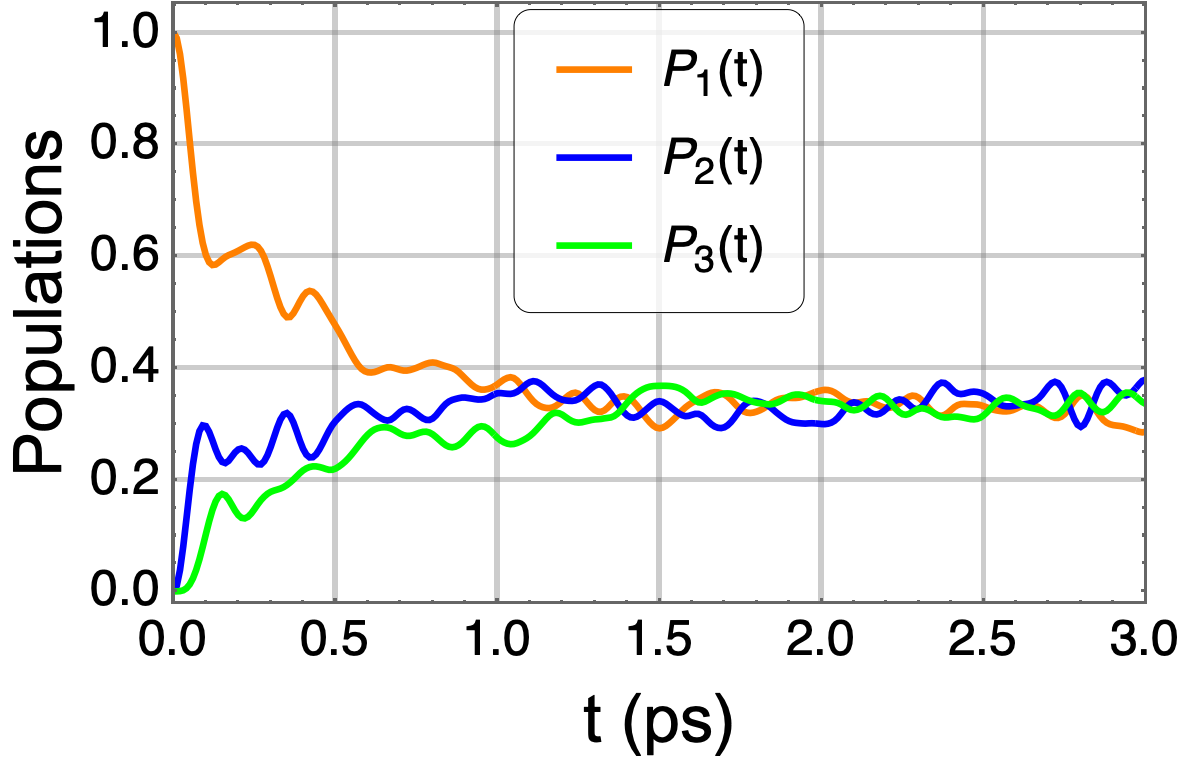}
    \caption{}
    \end{subfigure}
    %---------------------
       \centering
    \begin{subfigure}[t]{0.4\linewidth}
    \includegraphics[width=\linewidth]{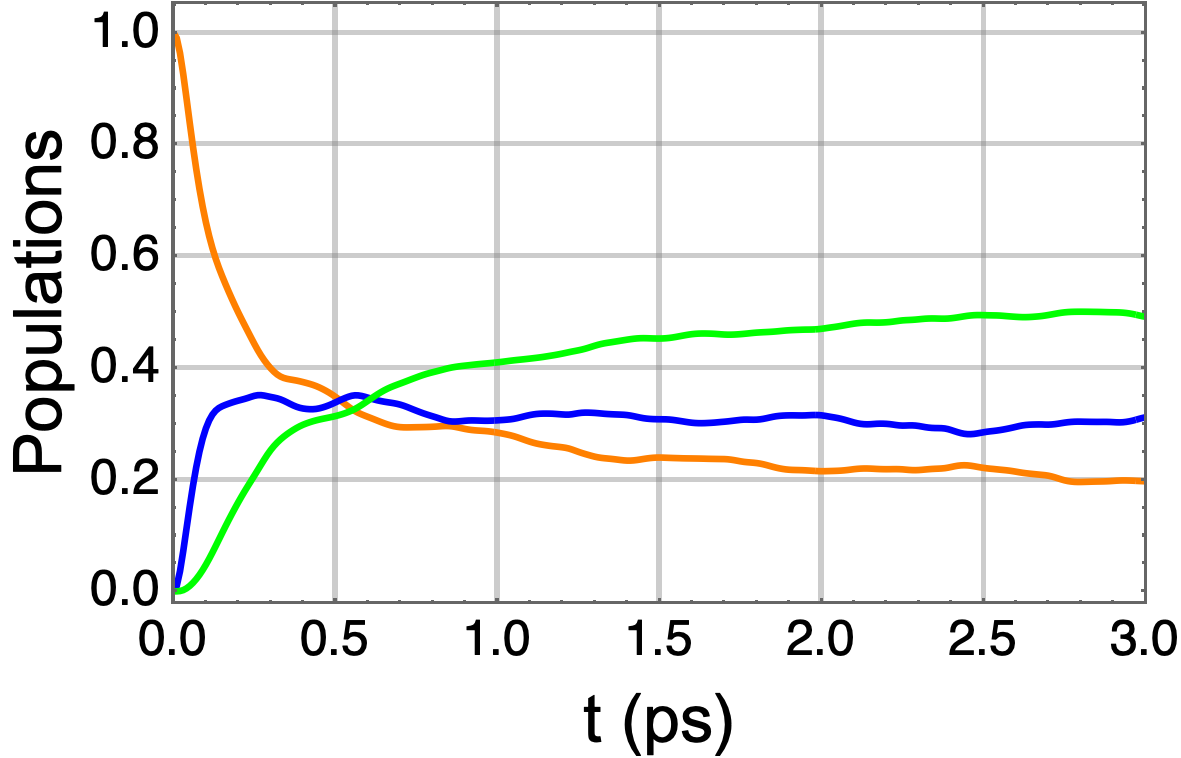}
    \caption{}
    \end{subfigure}
    %-------------------
    \centering
    \begin{subfigure}[t]{0.4\linewidth}
    \includegraphics[width=\linewidth]{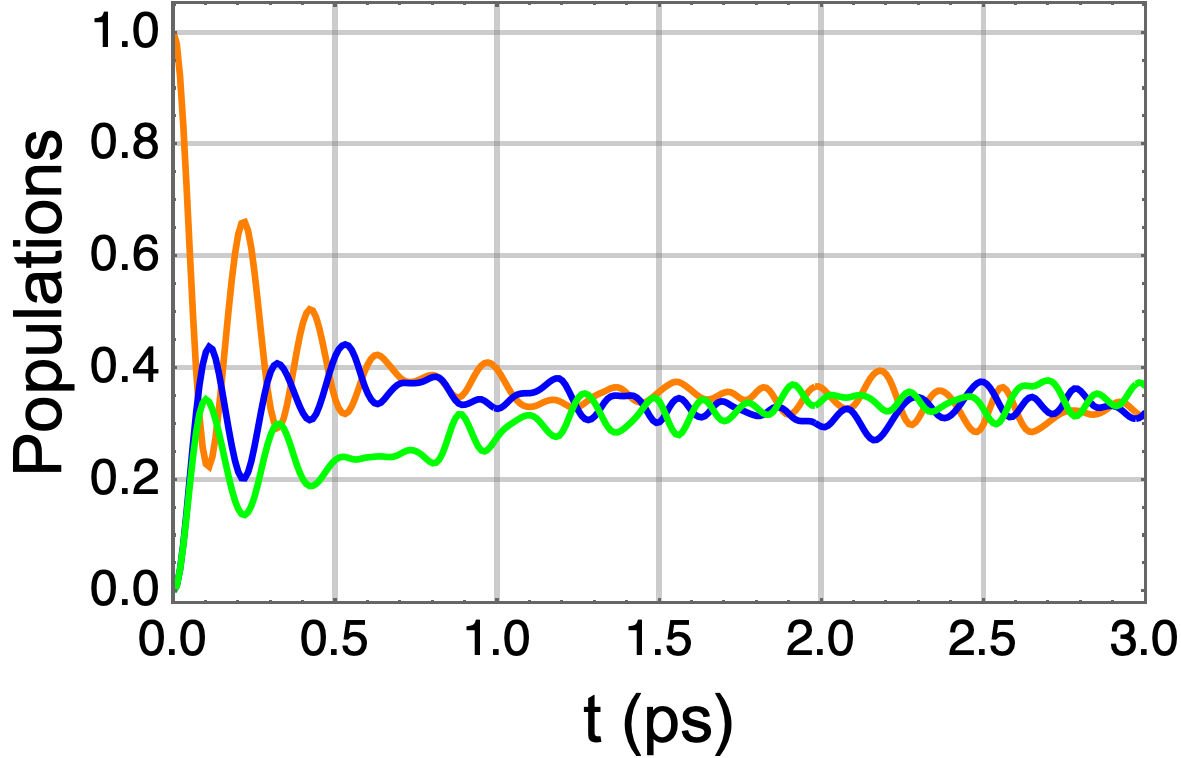}
    \caption{}
    \end{subfigure}
    %---------------------
       \centering
    \begin{subfigure}[t]{0.4\linewidth}
    \includegraphics[width=\linewidth]{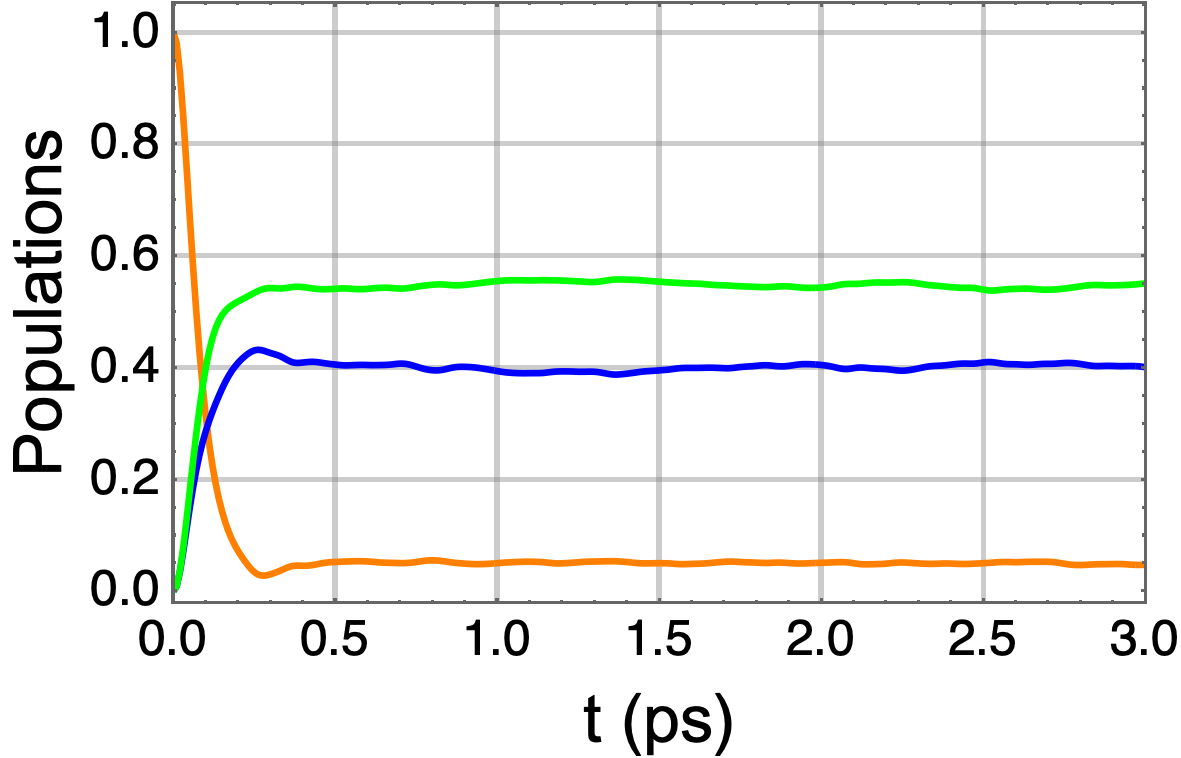}
    \caption{}
    \end{subfigure}
   \caption{The dynamics of populations for three-site system at noise amplitude $\Omega=14$ ps$^ {-1}$ and frequency $\nu=4$ ps$^ {-1}$ for linear chain (first row) and for the ring geometry (second row) for RTN only (a),(c) and bath+RTN models (b),(d). 
   }
    \label{fig:population dynamics}
\end{figure}

We consider two different environmental models -- RTN only and bath+RTN -- and the dynamics of the populations for the 3-site system
using both linear chain and ring configurations are given in Fig.~\ref{fig:population dynamics}. We use high noise frequency and intermediate noise amplitudes, and
initially the exciton is localized at site 1.
In Figure~\ref{fig:population dynamics}(a) and (c) we display the time evolution of the populations for the noise only model, which approaches the maximally mixed state for both the linear chain and ring geometry networks in the long time period, even if those populations become qualitatively different in the short time period. This result is expected due to the nature of RTN noise: it is symmetric, hence it corresponds to the infinite temperature~\cite{Gurvitz2017a}. However, the effect of the RTN on the system in contact with the thermal environment has two components. Since the site energies are changed in a time dependent way due to the external noise, the system exists in a non-equilibrium state. External noise, depending on its amplitude and frequency, might change the energetic order of the states of the system. This can be seen in Fig.~\ref{fig:population dynamics}(b) and (d), where it is found that the exciton is delocalized over all three sites with comparable populations for the linear chain. For the ring geometry, the steady-state is quite different; here the exciton is shared by the sites 2 and 3. Comparing the steady-state populations of site-3 for the RTN only and bath+RTN models, for both interaction geometries, one might deduce that the excitation transport is enhanced with simultaneous action of the quantum bath and the external noise. 

\begin{figure}[!hbt]
    \centering
    \begin{subfigure}[t]{0.4\linewidth}
    \includegraphics[width=\linewidth]{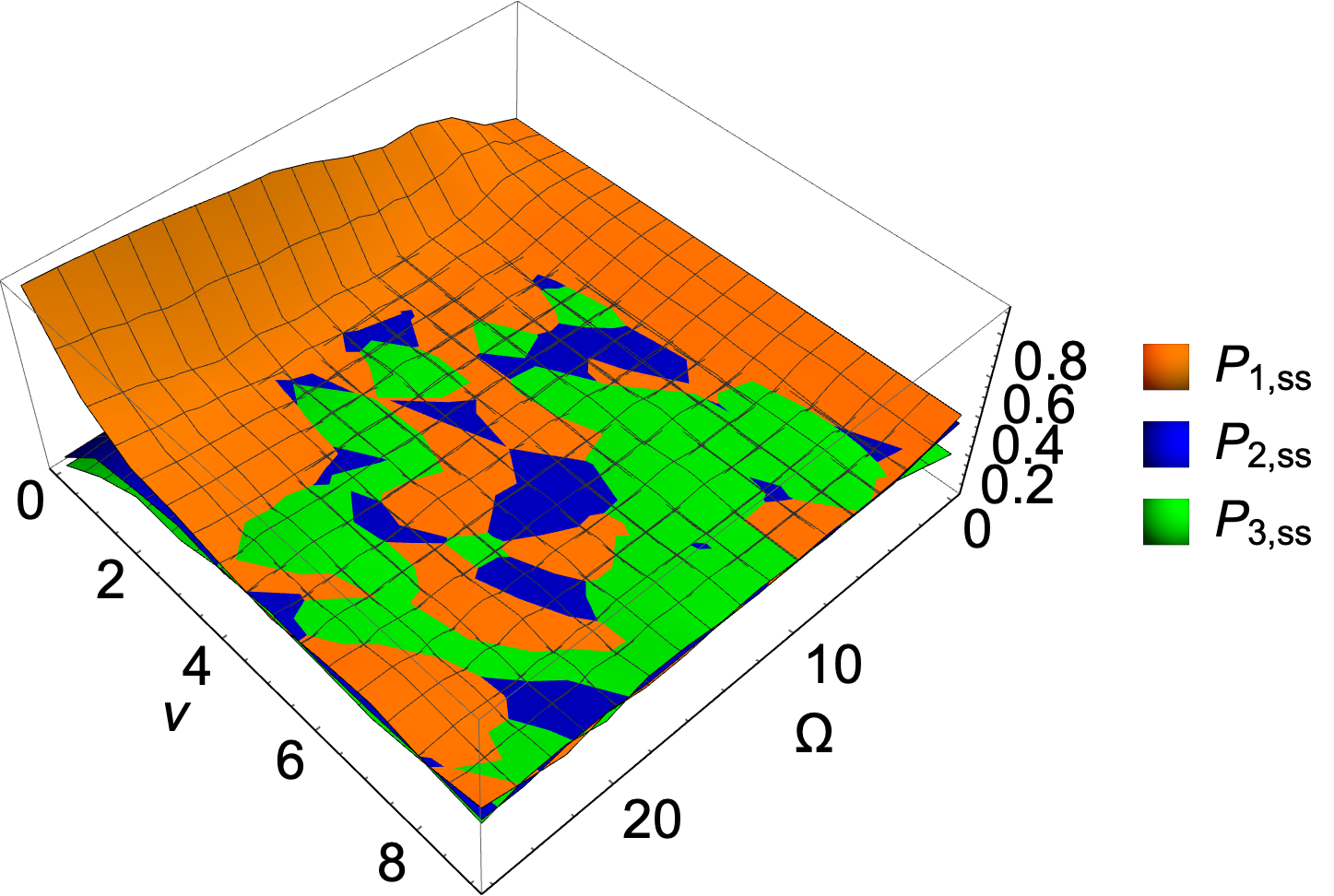}
    \caption{}
    \end{subfigure}
    %---------------------
       \centering
    \begin{subfigure}[t]{0.4\linewidth}
    \includegraphics[width=\linewidth]{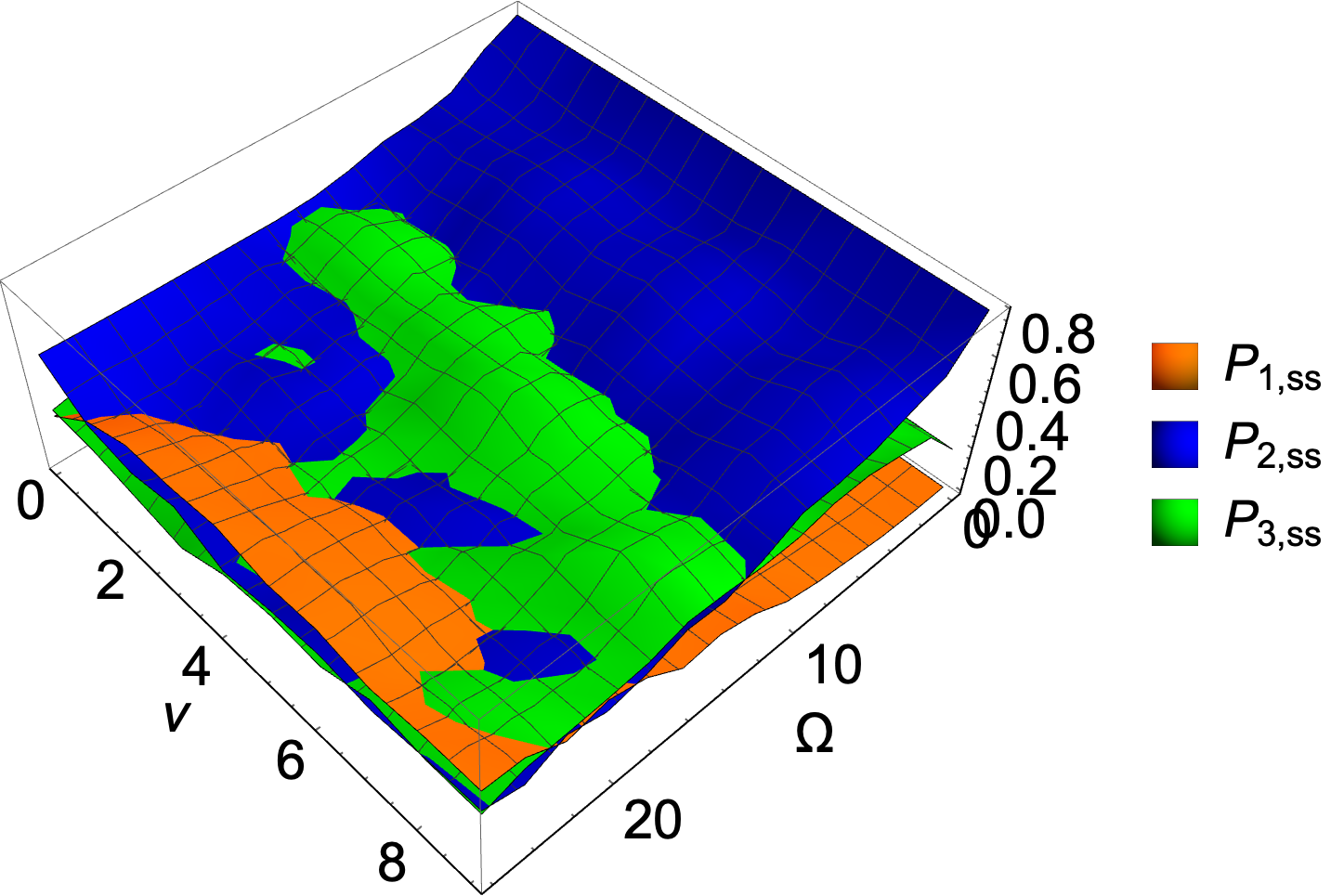}
    \caption{}
    \end{subfigure}
    %-------------------
    \centering
    \begin{subfigure}[t]{0.4\linewidth}
    \includegraphics[width=\linewidth]{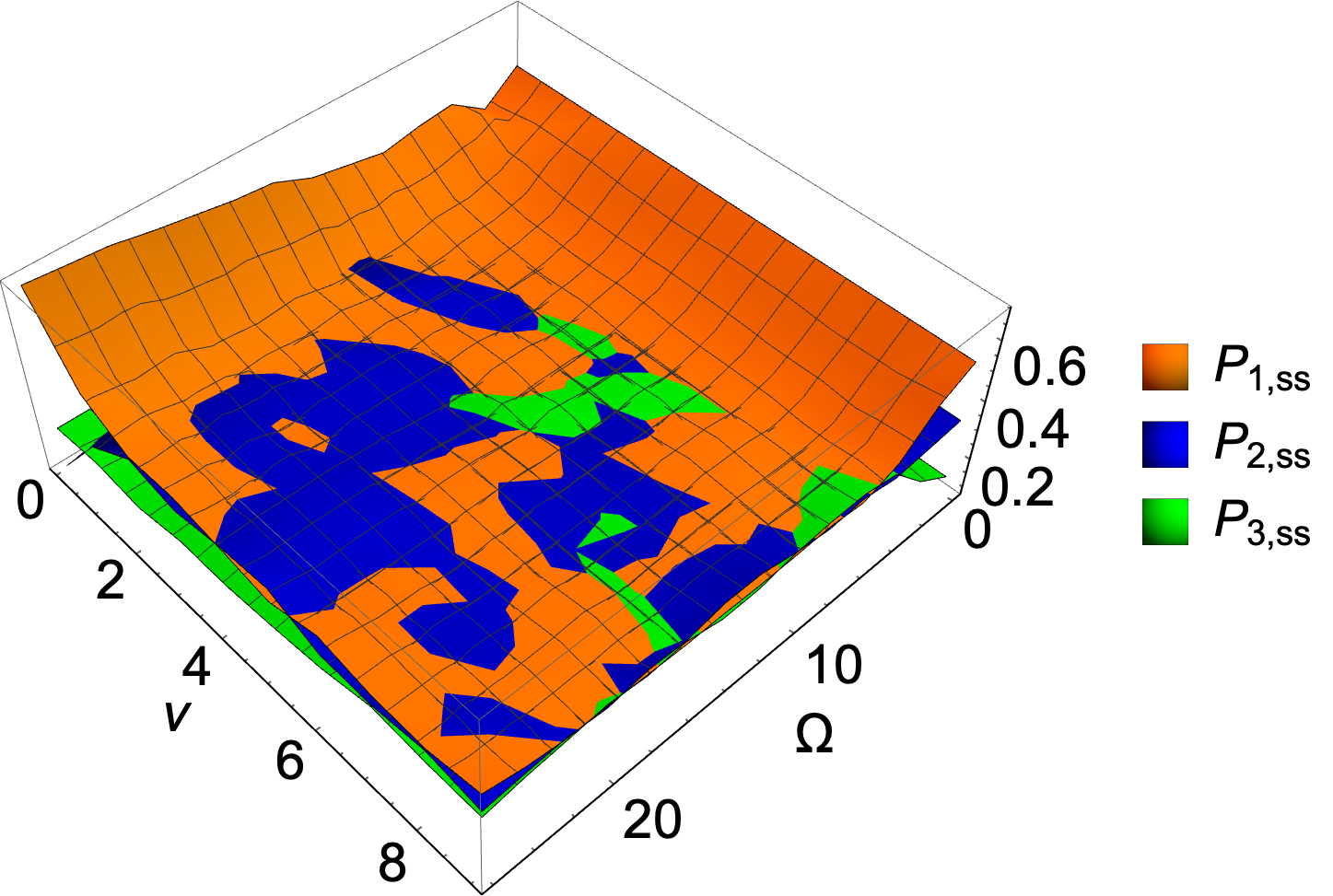}
    \caption{}
    \end{subfigure}
    %---------------------
       \centering
    \begin{subfigure}[t]{0.4\linewidth}
    \includegraphics[width=\linewidth]{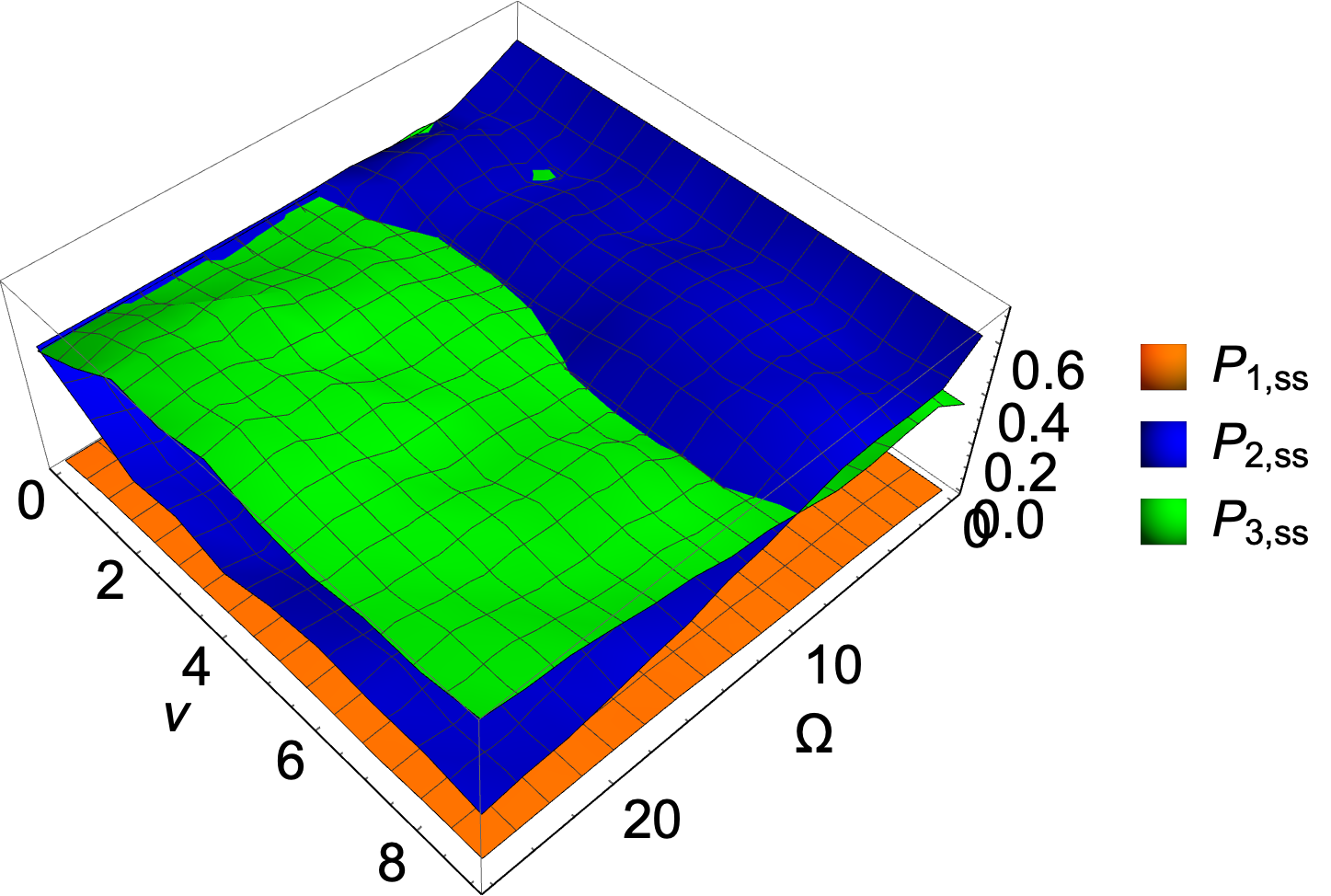}
    \caption{}
    \end{subfigure}
   \caption{Steady-state populations -- $P_{1,ss},P_{2,ss},P_{3,ss}$ --  as a function of noise frequency $\nu$ and noise amplitude $\Omega$ at $t=5$ ps for linear chain  in the first row and at $t=3$ ps for the ring geometry in the second row. RTN only model is displayed in (a),(c), while bath+RTN model is given in (b),(d).}
    \label{fig:steady state dynamics}
\end{figure}

To further explore the noise dependence of the system dynamics, in Fig.~\ref{fig:steady state dynamics} we display the steady-state populations as function of noise amplitude $\Omega$ and noise frequency $\nu$ for the RTN only and bath+RTN cases using again linear chain and ring geometries.  In the RTN only case [Fig.~\ref{fig:steady state dynamics} (a) and (c)], the steady-state populations tend to approach the maximally mixed state for both the linear chain and ring geometries at large $\nu$ and $\Omega$, which is similar to what we observed in Figs.\ref{fig:population dynamics}(a) and (c). Besides, it is clearly observed that at the low frequency limit (where the correlation time of the noise is larger than the dynamics time), the transport is prevented in the linear chain geometry in Fig.~\ref{fig:steady state dynamics} (a). This result stems from the fact that the anti-correlated noise increases the energy difference between the sites with increasing $\Omega$, which prevents the transfer of the exciton to the other sites above a certain noise amplitude.

On the other hand, when having thermal bath in addition of RTN, the steady-state populations have very interesting behavior depending on the site coupling configuration, see Figs.~\ref{fig:steady state dynamics}(b) and (d). Here, the $\Omega=0$ curve displays the steady-state populations for the system in contact with the thermal bosonic bath only -- this dynamics have been studied by different authors over the years~\cite{Plenio2008,Chin2010,Leggett87}. The system Hamiltonian in the variational frame without the external RTN indicates that the site-2 has the lowest energy in the thermal equilibrium (see Table~\ref{tab:par1}) and also the energy of the site-3 is close to that of site-2. This explains why the site-2 has initially ($\Omega=0$) the highest steady state population for the linear chain.

Moreover, we observe that for both the linear chain and ring configurations, the exciton is delocalized over site-2 and 3 with comparable probabilities when the external noise has absent or small amplitude, see Fig.~\ref{fig:steady state dynamics}(b) and (d). As the noise amplitude increases, Fig.~\ref{fig:steady state dynamics}(b) displays that $P_{1,ss},P_{2,ss}$ and $P_{3,ss}$ eventually relax to the maximally mixed state for the nearest-neighbor coupling network, as is the case for the RTN only model -- even if the site-3 has the maximum value at around $\Omega=10$. However, when going to ring configuration with coherently coupling also the  site-1 and site-3, one can see that the steady-state population of site-3 reaches the maximum value after crossing the resonance boundary of the bare system for the ring geometry in Fig.~\ref{fig:steady state dynamics}(d).
Understanding the interplay of coupling geometry (e.g. the additional pathways for interference, as well as different structure of the energy levels introduced by the ring geometry), thermal bath and random telegraph noise is a non-trivial task that is worth investigating in future work.
Thereby, we find that the site-3 is clearly the most populated one when having high amplitudes of the noise for the ring geometry and the model includes both the RTN and the bath.

In general, this means that RTN+bath model seems to be very efficient in terms of reaching the target site-3.
Does this mean that high occupation of site-3 also corresponds to the most efficient transport and shortest trapping time in our proposed model? To explore these questions and account for the excitation leaving the system, we add a sink to the target site-3. This is associated to having a reaction center where the energy is converted to chemical compounds. With all the above ingredients, we can then also see whether the transport can be enhanced even further when comparing to recent results of~\cite{Uchiyama2017a} and having also the action of the bath accounted for in addition of RTN.

%%%%%%%%%%%%%%%%%
\subsection{Transport efficiency and average trapping time}
\label{sub:efficiency}
We now investigate in detail the dependence of transport efficiency $\eta$ and average trapping time $\langle t\rangle$ in the above described three site dissipative 
systems including also the influence of sink site which transfer the exciton to reaction center. Here, 200 RTN realizations are used to average the reported quantities. 
The transport efficiency $\eta$ essentially quantifies which fraction of the population has been transferred to reaction center within a given interval of time.
The average trapping time $\langle t\rangle$, in turn, describes how fast is the transfer of the excitation to reaction center.
Subsequently, these two quantities can defined and expressed in terms of the site populations in the following way~\cite{Uchiyama2017a,Rebentrost2009,Cao2009}:
\begin{eqnarray}
    \langle t\rangle&=&\sum_{i}\int_{0}^{\infty}\rho_{ii}(t)\,dt,\quad
    \eta=2\kappa\int_{0}^{tu}\rho_{nn}(t)\,dt.
    \label{eq:avTrap_eff}
\end{eqnarray}
\noindent Here, $n=3$ indicates the trap site and we have chosen $\mathrm{tu}=7\,\mathrm{ps}$ following similar arguments as 
in~\cite{Uchiyama2017a} . The trapping rate $\kappa$ in Eq.~\eqref{eq:master} is chosen to have a relatively low 
value, 0.5 ps$^{-1}$, to be reasonably distant from the localization limit~\cite{Pelzer2013}.  

Notice that recently there has been a study properly accounting for the intensity of the incoming energy (sunlight) for realistic photosynthetic systems \cite{Gurvitz2019}. Here, however, we focus on engineered systems for excitation transport, therefore we employ the commonly used figures of merit defined above.

\begin{figure}[!hbt]
    \centering
    \begin{subfigure}[t]{0.4\linewidth}
    \includegraphics[width=\linewidth]{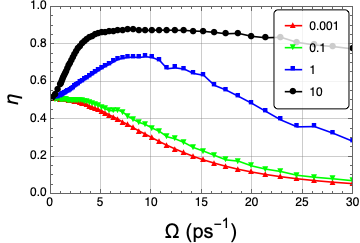}
    \caption{}
    \end{subfigure}
    %---------------------
       \centering
    \begin{subfigure}[t]{0.4\linewidth}
    \includegraphics[width=\linewidth]{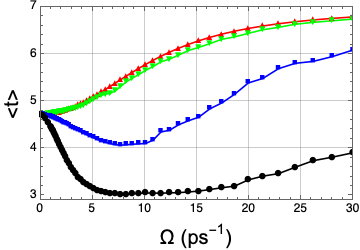}
    \caption{}
    \end{subfigure}
    %-------------------
    \centering
    \begin{subfigure}[t]{0.4\linewidth}
    \includegraphics[width=\linewidth]{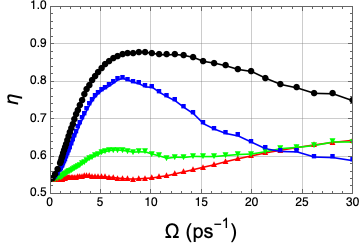}
    \caption{}
    \end{subfigure}
    %---------------------
       \centering
    \begin{subfigure}[t]{0.4\linewidth}
    \includegraphics[width=\linewidth]{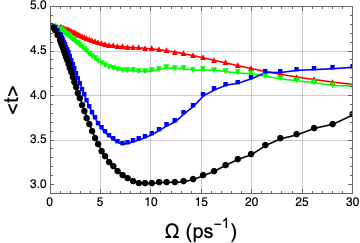}
    \caption{}
    \end{subfigure}
   \caption{ RTN only model. Transport  efficiency $\eta$ (a,c) and average trapping time $\langle t\rangle$ (b,d) as a function of noise amplitude $\Omega$ at different noise frequencies $\nu$ in strong-electronic regime with linear chain (a,b) and ring (c,d) configurations. }
    \label{fig:freq_NOnly}
\end{figure}

We first present the transport efficiency and average trapping time for the RTN only model, which neglects the thermal environment completely. The noise only model considered here is similar in spirit to the one studied by Chen and Silbey~\cite{Localization2011} and simulates the dynamics of the system $H_S(t)$ by ensemble averaging over different realizations of the RTN processes. Figure~\ref{fig:freq_NOnly} displays the behavior of $\eta$ and $\langle t\rangle$ as  function of noise amplitude $\Omega$ at various noise frequencies $\nu$ for linear chain and ring configurations. Efficiency plots for both the linear chain [Fig.~\ref{fig:freq_NOnly}(a)] and the ring [Fig.~\ref{fig:freq_NOnly}(c)] configurations show similar resonant enhancement of $\eta$ with increasing noise frequency $\nu$. The increase in transport efficiency with noise amplitude $\Omega$ is somewhat greater for the ring geometry compared to the one in the linear chain, as expected, probably due to higher number of decay channels in the former geometry. The enhancement is most pronounced when the noise amplitude $\Omega$ values are close to the site energy differences. 

The $\Omega$-dependence of $\eta$ is noticeably different for both geometries at very low noise frequencies compared to that at intermediate and fast ones. At very low noise frequencies, the noise does not flip in the considered time interval and the averaging over the two solutions of Eq.~(\ref{eq:master}) with $\alpha(t)=\pm 1$. For the linear chain in Fig.~\ref{fig:freq_NOnly}(a) the 
transport efficiency decreases to very low values with increasing noise amplitude indicating that high amplitude noise prevents the transport which is not observed 
for the ring geometry [see Fig.~\ref{fig:freq_NOnly}(c)]. The numerical trends in average trapping times [Figs.~\ref{fig:freq_NOnly}(b) and (d)] are almost opposite of those observed for the efficiencies; at those $\Omega$ values where $\eta$ is enhanced, $\langle t\rangle$ is strongly suppressed demonstrating fast transport of the excitation to the reaction center. This may be partially expected since efficient transport is coinciding here with fast transport - even though the two concepts are slightly different.
In gerenal, the results presented above show that a simple classical noise might enhance transport efficiency in a quantum setting similar to ENAQT phenomena.
\begin{figure}[!hbt]
    \centering
    \begin{subfigure}[t]{0.4\linewidth}
    \includegraphics[width=\linewidth]{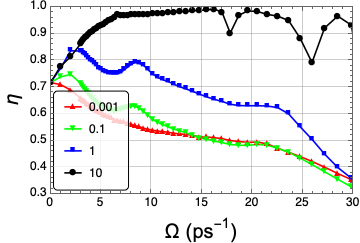}
    \caption{}
    \end{subfigure}
    %---------------------
       \centering
    \begin{subfigure}[t]{0.4\linewidth}
    \includegraphics[width=\linewidth]{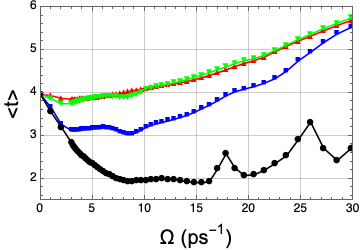}
    \caption{}
    \end{subfigure}
    %-------------------
    \centering
    \begin{subfigure}[t]{0.4\linewidth}
    \includegraphics[width=\linewidth]{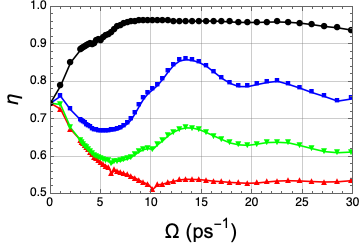}
    \caption{}
    \end{subfigure}
    %---------------------
       \centering
    \begin{subfigure}[t]{0.4\linewidth}
    \includegraphics[width=\linewidth]{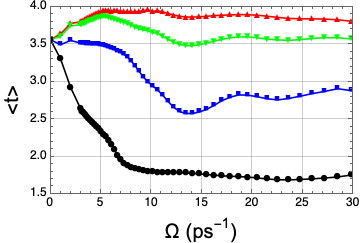}
    \caption{}
    \end{subfigure}
   \caption{Bath+RTN model. Transport  efficiency $\eta$ (a,c) and average trapping time $\langle t\rangle$ (b,d) as a function of noise amplitude $\Omega$ at different noise frequencies $\nu$ in strong-electronic regime with linear chain (a,b) and ring (c,d) configurations. 
   }
    \label{fig:freq_Bath+Noise}
\end{figure}

Let us turn our attention now to the bath+RTN case. Figure~\ref{fig:freq_Bath+Noise} shows the behavior of transport efficiency and average trapping time with the same RTN parameters as in Fig.~\ref{fig:freq_NOnly} but including now also the influence of the bath. Here, the $\Omega=0$ line indicates only a thermal bath model which does not contain the effects of the RTN  on the system -- this framework has been also studied by some authors~\cite{Plenio2008,Chin2010}. Without any external RTN noise, the transport efficiency 
has almost equal values between the two site configurations, see Figs.~\ref{fig:freq_NOnly} (a) and (c), and Figs.~\ref{fig:freq_Bath+Noise} (a) and (c) at the values $\Omega=0$. However, in the former case we have  $\eta\approx 0.50$ and in the latter case $\eta\approx 0.75$. This difference in transport efficiency 
 may be based on the presence of thermal fluctuations in the latter case which increases the probability of shifting the exciton to reach the trapping site.
 
In general, the set of results for RTN only, Fig.~\ref{fig:freq_NOnly}, and the
bath+RTN model, Fig.~\ref{fig:freq_Bath+Noise}, show that the dependence of both
the efficiency and trapping time on the noise parameters is qualitatively
different for different site configurations, and that the bath+RTN model has
very rich ENAQT behavior. 
Typically, simulations of the ENAQT in FMO has found that there is an optimal
region where the transport efficiency is enhanced by noise parameters depending
on the proposed model. However, we show that the behavior of transport
efficiency for RTN only and RTN+bath model is different from the results
obtained in both a thermal bath~\cite{Plenio2008,Chin2010} and the pure
dephasing~\cite{Montiel2014,Mohseni2014} cases.  That is, in the RTN-only model,
there is a single maximum in the dependence of the efficiency on the noise amplitude
and noise frequency depending on the site configurations. The structure is
richer for the bath+RTN model, where one can see resonance peaks.

It is also interesting to note that for both models the enhancement rate in the
transport efficiency $\eta$  in the linear chain configuration is relatively
higher than in the ring geometry for the parameter range explored here. The
enhancement rate is obtained by using the values at $\Omega=0$ and the overall
maximum value for any $\nu$ and $\Omega$.  For the linear chain, the rate is
about 80\% (for RTN only, 0.50 - 0.90 ) and 39\% (for bath+RTN, 0.72 - 1.00),
while for the ring geometry it is about 63\% (RTN only, 0.54 - 0.88) and 30\%
(bath+RTN, 0.74 - 0.96). The enhancement is more prominent for the noise only
model, which is expected because the presence of the quantum bath already helps
the transport. On the other hand, it can be seen clearly from
Fig.~\ref{fig:freq_NOnly} and \ref{fig:freq_Bath+Noise} that the noise frequency
$\nu$ dependence of $\eta$ is similar for both models; the transport efficiency
increases with the increasing $\nu$. This tendency is, however, not observed in
the results corresponding to the steady-state populations of the system [see
Fig.\ref{fig:steady state dynamics}]. This means that, although one could expect
that high steady-state populations equal indicate efficient transport to the
reaction centre~\cite{Montiel2014,Jesenko2013}, this expectation is not always
correct.   

\begin{figure}[!hbt]
    \centering
    \begin{subfigure}[t]{0.4\linewidth}
    \includegraphics[width=\linewidth]{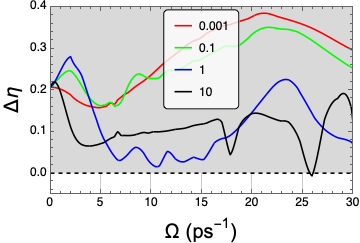}
    \caption{}
    \end{subfigure}
    %---------------------
       \centering
    \begin{subfigure}[t]{0.4\linewidth}
    \includegraphics[width=\linewidth]{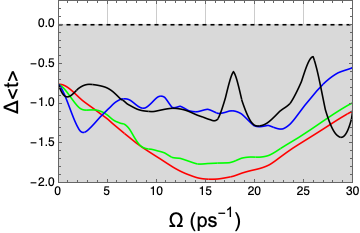}
    \caption{}
    \end{subfigure}
    %-------------------
    \centering
    \begin{subfigure}[t]{0.4\linewidth}
    \includegraphics[width=\linewidth]{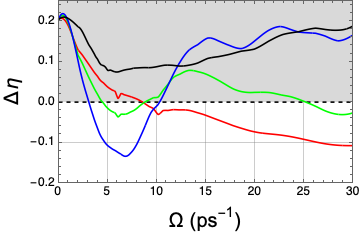}
    \caption{}
    \end{subfigure}
    %---------------------
       \centering
    \begin{subfigure}[t]{0.4\linewidth}
    \includegraphics[width=\linewidth]{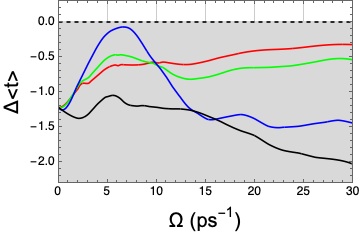}
    \caption{}
    \end{subfigure}
   \caption{
   The difference of efficiency $\Delta\eta$ (a),(c) and average trapping time $\Delta \langle t \rangle$ (b),(d) 
   between bath+RTN and RTN only models for the 3-site system  as a function of
   noise amplitude at different noise frequencies -- for the linear chain in the
   first row and for the ring geometry in the second row. Note that the grey
   shades indicate where the bath+RTN model provides superior performance
   compared to RTN only model.
   }
    \label{fig:difference_eff_avt}
\end{figure}

In order to show that the bath+RTN model significantly improves the efficiency and transport time compared to RTN only model, we display the differences
$\Delta\eta$ and $\Delta\langle t\rangle$ between the former and the latter in Fig.~\ref{fig:difference_eff_avt}.  The results show that the bath+RTN model is constantly dominant against the RTN only model for linear chain in terms of both measures, see Figs.~\ref{fig:difference_eff_avt}(a) and (b) and the corresponding gray colored areas.
For ring configuration in Figs.~\ref{fig:difference_eff_avt}(c) and (d), the bath+RTN model is superior for the average trapping time [Figs.~\ref{fig:difference_eff_avt}(d)] while for efficiency there are specific limited areas of noise amplitude and frequency, where RTN only dominates. 
In general, we conclude that when looking for efficient and fast transport, having the influence of both classical RTN noise and quantum bath provides the most useful avenue.

%In Fig.~\ref{fig:difference_eff_avt}, in order to show the region where bath+RTN model has significant effect on both efficiency and averaging trapping time, we plot the difference of both $\Delta\eta$ and $\Delta\langle t\rangle$ between bath+RTN and RTN only models versus noise amplitudes at various noise frequencies for linear chain and ring geometry. We show that bath+RTN model is constantly dominant against the RTN only model for linear chain in terms of both measures in figure~\ref{fig:difference_eff_avt}(a) and (b), while for ring geometry there is a region where the noise only model has relatively efficient than the bath+RTN model at low and intermediate frequencies in figure~\ref{fig:difference_eff_avt}(c) and (d). The results we observed in the set of Fig.~\ref{fig:difference_eff_avt} is already supported by the previous ones.

\section{Conclusions}
\label{sec:conclusion}
In this work, we have developed a formalism for and considered multi-site dissipative systems influenced by both the classical RTN noise and quantum baths.
Technically, this requires the use of the variational polaron transformation for the derivation of the corresponding master equation, generating stochastic realizations of open system density matrix evolutions and using the ensemble averaging over noise realizations to obtain the final result for multi-site density matrix.

We used the method to study the efficiency of the energy transport in a three-site system by focussing and comparing the results from two different models -- RTN only and bath+RTN model, where RTN contains full (anti)correlations between the three sites.  One of the motivations was to go beyond a recent observation that correlated stochastic noise improves the transport efficiency~\cite{Uchiyama2017a}. The results for steady state populations, by using a chain and ring configurations, clearly demonstrate in both cases that there exists a considerable parameter region where the target site population is significantly increased when combining stochastic noise with a quantum bath. 
To take into account the transfer of an excitation from the target state to the
reaction center, we consider similar models as above but now with a sink, and
calculate the transport efficiency and average trapping time -- which are
commonly used to quantify the efficiency of the transport~\cite{Uchiyama2017a}.
Here, the ENAQT behavior has quite a rich structure in terms of the frequency and amplitude of the random noise. The results clearly show that in almost all 
regions of the used parameter space, the combined influence of bath+RTN is superior to RTN only model when trying to achieve efficient and fast transport.
However, many questions remain open for future studies. How does the efficiency change when increasing the number of sites and the complexity of the coupling configurations between the sites within our framework? What is the role of changing the temperature of the environment and the spectral character of the local baths of the sites?
In general, our result opens the avenue to study, e.g., the questions above and when looking for ways and characterization of efficient
 energy transport in engineered and ambient systems.

\ack

We acknowledge fruitful discussions with Matteo G. A. Paris and Resul Eryi\u{g}it. AK acknowledges financial support from The Scientific and Technological research Council of Turkey (TUBITAK) Project no.
1059B191700098. MR and JP acknowledge
financial support from the Academy of Finland via the
Centre of Excellence program (Project no. 312058 as well as Project no. 287750).
MR acknowledges support from the Turku Collegium for Science and Medicine. 

\newpage
\providecommand{\newblock}{}

\end{document}